\documentclass[aps,reprint,amsmath,amssymb,showkeys,twocolumn]{revtex4-2}
\usepackage{graphicx}
\usepackage{amscd}
\usepackage{dcolumn}
\usepackage{bm}

\begin{document}

\title{Barrow entropy and spacetime foam}


\author{Yu.L. Bolotin}
\affiliation{National Science Center Kharkov Institute of Physics and Technology, 1 Akademicheskaya str.,  Kharkov, 61108, Ukraine}
\author{V.V.Yanovsky}
\email{yanovsky@isc.kharkov.ua}
\affiliation{V. N. Karazin Kharkiv National University, 4 Svobody Sq., Kharkiv, 61022, Ukraine}
\affiliation{Institute for Single Crystals, NAS Ukraine, 60 Nauky Ave., Kharkov, 31001, Ukraine}

\date{\today}

\begin{abstract}
Quantum gravitational effects, on the one hand, lead to a limitation in the accuracy of measuring spatial and time intervals, and, on the other hand, they generate a discrete of spacetime structure (quantum foam). The common source of both measurement limitations and discreteness of space-time are quantum fluctuations, so their characteristics must be related to each other. We study such a relationship using Barrow entropy as a source of fractal space-time structure. The minimum inaccuracy in measuring space-time intervals is expressed through the Barrow entropy parameter. The connection between the level of fractality and the speed of information processing is considered.
\end{abstract}


\maketitle

\section{Introduction}

Planck was the first to propose combining the fundamental constants  $c, \hbar$, which are independent of dynamics, with a constant $G$  that describes the gravitational interaction in the nonrelativistic limit. This made it possible to construct a fundamental unit of length  $l_p =\sqrt{\frac{\hbar G}{c^3}}$.

There are two ways to include the fundamental length in the structure of space-time and, accordingly, two types of physical models \cite{divakaran2024matter,Hossenfelder_2013}.

The currently dominant approach treats space-time as continuous (a pseudo-Riemannian manifold), but does not allow physical measurements of lengths below some scale, much smaller than any scale currently available. In models of this type, the Planck length plays  role of the fundamental length.
An alternative approach is to consider space-time to be inherently fundamentally discrete. The implementation of such an approach, even the classical level, is a difficult task, requiring a complete restructuring of the mathematical apparatus \cite{Loll_1998,Williams_2006,bolotin2020physics}.

As an intermediate stage of the transition from continuous to discrete space-time, we can consider the transformation of smooth surfaces into fractal ones \cite{mandelbrot1983fractal,BTY16}. Even on the smallest scales currently observable, spacetime appears smooth and structureless. However, the opinion that continuous space-time is just an idealization is gradually winning out. One of the main reasons forcing us to abandon the concept of continuum is the existence of quantum fluctuations. It is natural to assume that quantum effects  on Planck scales transform continuous space-time into a complex (turbulent) structure, called space-time foam (quantum foam) \cite{dewitt1964relativity}.

This restructuring of space-time structure is especially important in the vicinity of the black hole event horizon or cosmological horizons. The surface area of the horizon, which obviously changes during such a restructuring, determines the dynamics and thermodynamics of the object limited by the horizon.

\section{Measurement limitations}

Taking into account quantum gravitational effects requires a transition from the Heisenberg uncertainty principle to the generalized uncertainty principle (GUP) \cite{Veneziano_1986,Amati_1988tn,Witten_1996,Adler_2001}.

According to GUP, uncertainty in position $\Delta x$ and momentum in $\Delta p$ one dimension are related by
	 \begin{equation}\label{eq1}
\Delta x \geq	\frac{\hbar}{ \Delta p}	+l_P^2 \frac{\Delta p}{\hbar} 
	 \end{equation}
The dramatic consequence of taking gravity into account in the measurement process is the emergence of such a fundamental concept as the minimum length, $\Delta x_{min} \approx l_P$. The physics of this parameter is extremely simple. At small photon momenta (large wavelengths the localization of the measured object will be bad. For large photon momenta, its gravitational interaction with the measured object will again negatively affect the process of measurements. Between these two extremes, we can choose the photon momentum, which optimizes the measurement process.

The existence of a minimum length and, as a consequence, the discrete structure of space is, on the one hand, an almost unlimited source of new effects that are absent in the spatial continuum, and, on the other hand, it allows us to discover previously unknown connections between already studied phenomena. In addition, any theory with a minimum length inevitably contradicts a number of traditional formulations of both quantum mechanics and general relativity.

Taking into account quantum gravitational effects leads to a fundamental limitation on the accuracy of measuring $\delta l$ length $l$, \cite{Karolyhazy:1966zz}
\begin{equation}\label{eq2}
	\delta l \geq \left( l l_P^2\right)^{1/3}
\end{equation}
We emphasize that restriction (\ref{eq2}) is universal and does not depend on the method of measuring length.
Relations (\ref{eq2}) can be written in terms of uncertainty $\delta t=\frac{\delta l}{c}$ arising when measuring the time interval $t=\frac{l}{c}$
	 \begin{equation}\label{eq3}
		 \delta t \geq \left( t t_P^2 \right)^{1/3}
	 \end{equation}
	
	\section{The holographic principle and dynamic lattice}
		
The dependence of the accuracy of length measurement on the characteristic size of the measured object (\ref{eq2}) is equivalent to the holographic principle \cite{hooft2009dimensional,Susskind_1995}.
	
The simplest formulation of the holographic principle contains two statements. Firstly, all information contained in a certain region of space can be recorded (presented) on the border of this region, (a holographic screen). Secondly, the theory at the boundary of the studied region of space should contain no more than one degree of freedom per Planck area. In other words, the total number of degrees of freedom N obeys the inequality
	 	 \begin{equation}\label{eq4}
		 N\leq \frac{A}{l_P^2}
	 \end{equation}
	where $A$ is the area of the horizon.
\begin{figure}
	\includegraphics[width=4 cm]{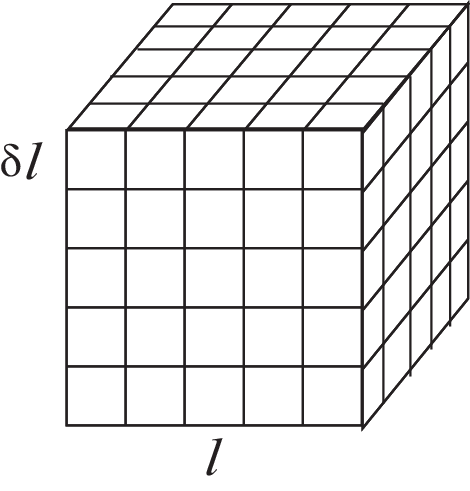}
	\caption{The division of volume into unit cells.}
	\label{fg1}
\end{figure}
	
The holographic principle as a strict statement is valid only for black holes. In other cases, it is just a hypothesis. Therefore, it is of interest to answer the question: is the holographic principle valid in a theory that combines quantum mechanics and gravity? On the one hand, such a theory generates a minimum length and provides a natural transition to a 3-dimensional spatial lattice. But, on the other hand, for the validity of the holographic principle it is required that all volumetric information can be recorded on a 2-dimensional surface lattice. Let's make sure that the inclusion of gravitational effects provides effective 2-dimensionality \cite{Ng_2003,PhysRevD.83.084003}. To do this, consider some 3-dimensional volume with a characteristic linear size $l$. In a theory with a minimum length  $l_{min} \approx l_P$, it is natural to assume that the volume is divided into cells of size  $l_P^3$. However, such a conclusion is erroneous. Due to a fundamental limitation  $\delta l \geq (l l_P^2)^{1/3}$, the minimum edge of a cell into which a cube with an edge $l$ can be divided is (see Fig. \ref{fg1})

Thus, spacetime foam leads to the formation of a dynamic lattice with a period depending on the size of the lattice. Therefore, the maximum number of cells in such a lattice
\begin{equation}\label{eq5}
	N=\left[ \frac{l}{\left(l l_P^2\right)^{1/3}}\right]^{1/3}=\frac{l^2}{l_P^2}
\end{equation}
If we associate one degree of freedom of the system with each cell, then the resulting relation reproduces the holographic principle.
The connection between the ``quantization'' of the horizon area and the holographic principle is directly determined by the evolution of horizon. Let a mass $\delta m$ be thrown over the horizon. It is clear that this mass should be enough to increase the area of the horizon by the Planck unit area. Then
	 	 \begin{equation}\label{eq6}
		 l_P^2 \leq \left(\frac{2 G}{c^2} \right)^2 \left[ (m+\delta m)^2 -m^2 \right]
	 \end{equation}
	It is convenient to replace the Planck length with the Planck mass. Then inequality (\ref{eq6}) will take the form
	 			\begin{equation}\label{eq7}
			\delta m^2 +2 m \delta m -m_P^2 \geq 0
		\end{equation}
This, in turn, means that the radius of the black hole is discrete
	 	\begin{equation}\label{eq8}
			r_{g,n+1}=\sqrt{r_{g,n}^2 +r_{g,P}^2}
		\end{equation}
where $r_{g,P}=\frac{2G m_P}{c^2}$ is the quantum of change in the radius of the black hole. Considering this relation is a recurrence equation with respect to the variable $n$, you can look for its solution
	\[ r_{g, n}=\sqrt{const + n r_{g,P}^2} \]
where $const$ is an arbitrary constant that determines the minimum radius of the black hole. If $const=0$ or $n \gg 1$ then
	 	\begin{equation}\label{eq9}
			r_{g, n} \propto \sqrt{n}
		\end{equation}
This result actually reproduces the holographic principle in the form (\ref{eq5}).

\section{Barrow entropy and spacetime foam}

Barrow \cite{Barrow_2020} proposed a phenomenological model of the formation of the fractal structure of the horizon for a stationary black hole. Barrow's reasoning was as follows. The surface area of a Schwarzschild black hole $A_g =4 \pi r_g^2$ determines its the entropy  $S= A_g \frac{c^3}{4 G} \approx \frac{A_g}{A_{Pl}}$. The entropy is proportional to the number of Planck squares  on the surface of the horizon. Consequently, quantum gravitational effects leading to the fractal structure of the horizon can significantly increase its entropy compared to the standard Bekenstein entropy. Of course, the increase in the surface area of a black hole during its evolution is consistent with the second law of thermodynamics.

Abstracting from the specific model of the black hole horizon, the area of the fractal surface of the horizon as a function of the gravitational radius $r_g$ behaves as $r_g^{2+\Delta}$, where  $0 \leq \Delta \leq 1$. The transition $(\Delta =0) \to (\Delta =1)$ realizes a transformation from a smooth 2-dimensional surface to  3--dimensional volume due to quantum deformation of the horizon surface. In other words, $\Delta =0$ it corresponds to the simplest structure of the horizon, and $\Delta =1$ corresponds to the greatest complexity of the horizon surface.

The fractality of the horizon surface generates the corresponding entropy (Barrow entropy)
\begin{equation}\label{eq10}
	S_B =\left(\frac{A}{4 A_P} \right)^{1+\frac{\Delta}{2}}
\end{equation}
where $A$ is the surface area of the horizon, $A_P$ is the Planck area, and $\Delta$ is a free parameter of the model $0 \leq \Delta \leq 1$. In the undeformed case $(\Delta =0)$, the Barrow entropy coincides with the Bekenstein entropy, and  the value $\Delta =1$ corresponding to the maximum deformation, leading to an increase in the fractal dimension of the horizon surface by one
	 \begin{equation}\label{eq11}
		 S \propto \frac{A}{A_{Pl}} \to S \propto \frac{A^{\frac{3}{2}}}{A_{Pl}^{\frac{3}{2}}}
	 \end{equation}
It seems natural to connect the degree of fractality of the horizon surface (parameter $\Delta$) with the period of the holographic lattice $\delta l$, determined by relation (\ref{eq2}). Such a connection must exist, since both the fractal structure and space-time discreteness are generated by quantum gravitational effects.

In any effective quantum field theory defined in a spatial region with a characteristic size $L$ and using ultraviolet cutoff $\Lambda$, entropy  $S \propto L^3 \Lambda^3$. If we assume that the entropy of any object of linear size $l$ in such a theory should be less than the entropy of a black hole of the same size, then
\begin{equation}\label{eq12}
	S\leq S_{BH} \approx \left( \frac{l}{l_P}\right)^2
\end{equation}
It seems natural to identify the inverse scale of UV cutoff $\Lambda^{-1}$ with a limiting limitation on the accuracy of length measurements . Then the relation (\ref{eq12}) take the form
\begin{equation}\label{eq13}
	\frac{l^3}{\delta l^3}\leq \left( \frac{l}{l_P}\right)^2
\end{equation}
Let us perform the same procedure for the Barrow entropy
\begin{equation}\label{eq14}
	\frac{l^3}{\delta l^3} \leq \left( \frac{l}{l_P}\right)^{2+\Delta}
\end{equation}
From here, we will find generalization of the fundamental inequality (\ref{eq2}) for the   Barrow entropy
\begin{equation}\label{eq15}
	\delta l \geq \left(l^{1-\Delta} l_P^{2+\Delta}\right)^{1/3}
\end{equation}
Similarly, relation (3) is transformed into
\begin{equation}\label{eq16}
	\delta t \geq \left(t^{1-\Delta} t_P^{2+\Delta} \right)^{1/3}
\end{equation}
Let us consider limiting cases for these relations. In the case $\Delta =0$  we reproduce fundamental inequalities (\ref{eq2}), (\ref{eq3}). Moreover, in the case $\Delta =1$, the accuracy of measuring space-time intervals ceases to depend on the size of the measured interval, $\delta l \delta t \geq l_P t_P$.

It is convenient to go to dimensionless variables
\begin{equation}\label{eq17}
	L \equiv \frac{l}{l_P} , \, \delta\bar{ l}\equiv \frac{\delta l}{l_P}
\end{equation}

\begin{equation}\label{eq18}
	T \equiv \frac{t}{t_P} , \, \delta \bar{t} \equiv \frac{\delta t}{ t_P}
\end{equation}
In these variables, inequalities (\ref{eq15}), (\ref{eq16}) take the form
\begin{equation}\label{eq19}
	\delta \bar{l} \geq \left( L^{1-\Delta}\right)^{1/3}
\end{equation}
\begin{equation}\label{eq20}
	\delta \bar{t} \geq \left(T^{1-\Delta} \right)^{1/3}
\end{equation}
The physical meaning of the obtained limit relations is simple. For $\Delta =0$  we are dealing with a smooth surface of the horizon and, as a consequence, we reproduce the inequalities (\ref{eq2}), (\ref{eq3}) generated by the Bekenstein entropy. For the case  $\Delta =1$ (maximum horizon deformation) $\delta l \delta t$  is transformed into the minimal size of self-similarity $l_P t_P$.

Barrow's concept of entropy can be extended to cosmological horizons by introducing holographic dark energy with density $\rho \sim H^{2-\Delta}$ \cite{Saridakis_2020b,Saridakis_2020,Mamon_2021,adhikary2021barrow,SarBas2021,Di_Gennaro_2022}. In this case, the parameter $\Delta$ can be expressed through the current values of cosmographic parameters $q_0 j_0$ \cite{bolotin2024note,bolotin2024cosmology,BolotinCSW_2022}
\begin{equation}\label{eq21}
	\Delta =2 \left[ \frac{(1-j_0)}{(1+q_0) (1-2 q_0)} +1\right]
\end{equation}
where $q$ and $j$ are deceleration and jerk parameters.
\begin{figure}
	\includegraphics[width=6 cm]{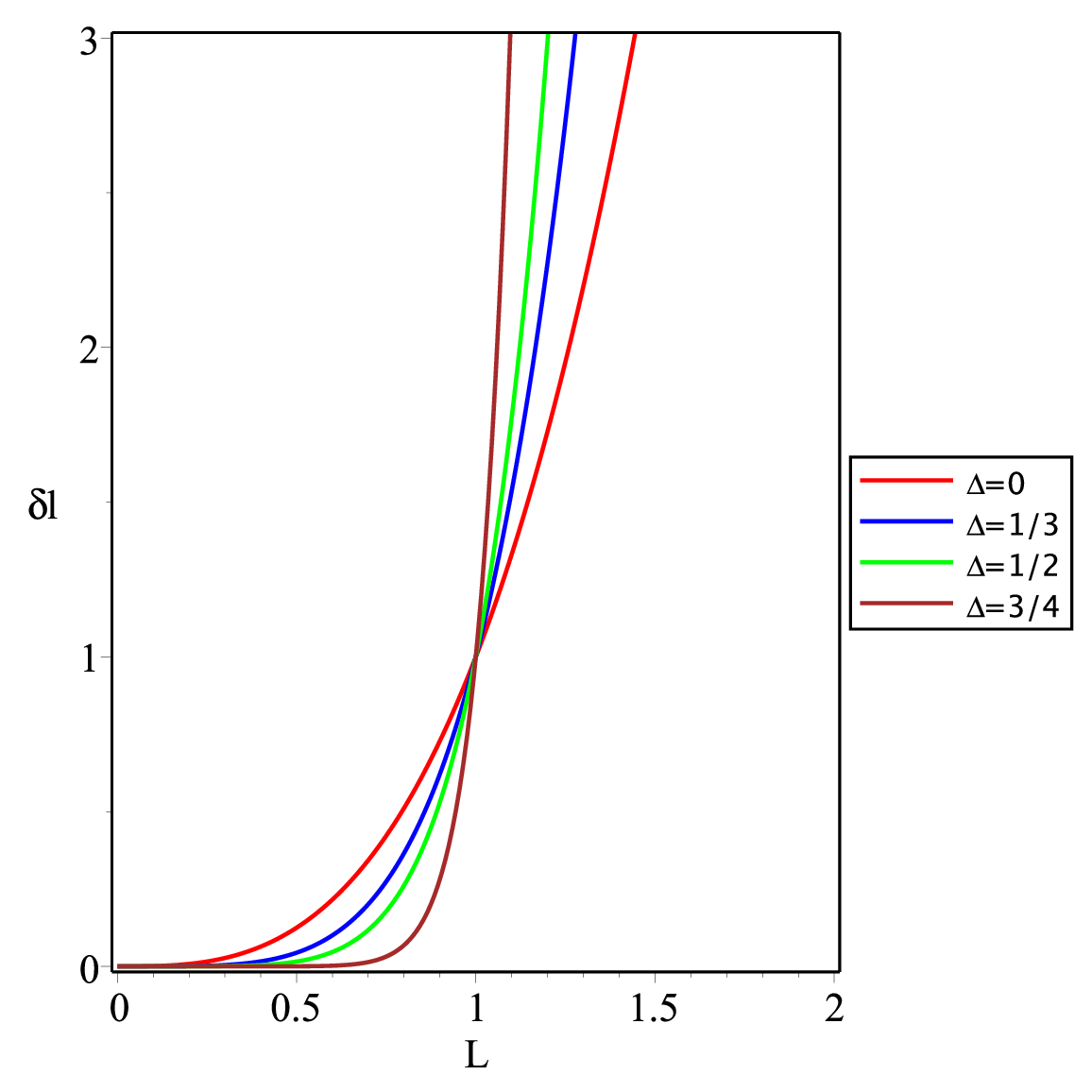}\\
	\caption{Measurement accuracy as a function of the maximum duration of the time interval.}
	\label{fg2}
\end{figure}

Note that in SCM $j=1$, therefore  $\Delta=2$, that goes beyond the permissible values  $\Delta$. In SCM, dark energy is realized in the form of a cosmological constant, and is naturally $\rho_e =CH^{2-\Delta}$  for  $\Delta =2$ transformed into a constant. In this case entropy (it cannot be called Barrow entropy, for which  $0 \leq \Delta \leq 1$) $S_B =\left( \frac{A}{A_0}\right)^{1+\frac{\Delta}{2}}$ \cite{PhysRevD123516,Komatsu_2014}. Note that the Barrow entropy represents a special case of the 4-parameter construction that can reproduce the range of entropies used for horizon thermodynamics \cite{Nojiri_2022n,nojiri2023microscopic}.

\section{Application examples}

As Barrow noted \cite{Barrow_2020}, based on an analysis of the process of evaporation of black holes, the deformation of the horizon by quantum fluctuations should accelerate the evaporation process. We will consider this statement from a more general point of view.In its meaning, a time interval measured with a given accuracy represents the time of correct functioning of a clock or, in other words, its lifetime. Therefore, using inequality (\ref{eq3}) we come to a far from obvious conclusion: the more accurate the clock, the less time it operates with a given accuracy. Let's test this concept using a black hole with Bekenstein entropy ($\Delta =0$) as a clock. Then the accuracy  $\delta t$  of such a device naturally choose  $r_g /c$. Now, using the upper limit of inequality (\ref{eq3}), we determine the lifetime of such a device (i.e., black hole)
\begin{equation}\label{eq22}
	T_{BH}=\left( \frac{r_g}{c} t_P^{-2/3}\right)^3 =\frac{8 m^3 G^2}{\hbar c^4}
\end{equation}
The resulting lifetime of a black hole in this unusual way coincides with the evaporation time of a black hole, which Hawking calculated. The only difference is in the numerical factor. Turning to Barrow entropy, we note that an increase in the degree of fractality (an increase in the parameter $\Delta$) leads to a decrease $\delta t$ ( increase in measurement accuracy). This, in turn, due to inequality (\ref{eq20}), will reduce the time interval available for measurement, i.e. lifetime of the object (see Fig. 2).

Increasing the speed of information processing and the volume of its storage is a key problem in both fundamental and applied physics. The prospect of practical implementation of quantum computing opens up new possibilities. It is obvious that the information characteristics of holographic models (storage volumes and processing speed) depend on the geometric structure of the horizon (holographic screen). As we saw above, quantum fluctuations, which lead to discrete space, play an important role far beyond the Planck scale. So for the observable Universe ($l \sim 10^{28}$cm) $\delta l \sim \left(l l_P^2 \right)^{1/3} \sim 10^{-13}$cm.  It is an almost ``macroscopic'' value of the order of the characteristic nuclear sizes. 

The existing physical limitations on the concentration of energy lead to a direct ban on the processes of excessive concentration of information. These restrictions can be obtained from the fundamental inequality (\ref{eq3}) for the Bekenstein entropy and inequality (\ref{eq20}) in the presence of a fractal structure of the horizon. Let us apply inequality (\ref{eq20}) to assess the efficiency of the computer. Quantity $\nu =\delta t$ can be interpreted as the operating frequency, and  $I=\frac{t}{\delta t} =\frac{T}{\delta \bar{t}}$  represents the maximum amount stages of information processing. Then inequality (\ref{eq20}) is transformed into
\begin{equation}\label{eq23}
	I^{1-\Delta} \nu^{2+\Delta}\leq t_P^{-(2+\Delta)}
\end{equation}
For $\Delta =0$ inequality (\ref{eq23}) reproduces the well-known \cite{Ng_2003} constraint.

\section{Conclusion}

Barrow entropy model can be thought of as a way of generating fractal structure in the form of space-time foam. We have shown that the period of the dynamic lattice simulating this quantum foam directly depends on the only model parameter of Barrow entropy and have found this dependence. Limiting versions of this dependence were analyzed, corresponding to the undeformed horizon surface and the maximum deformed surface by quantum fluctuations. The inequality is used to estimate the lifetime of an object operating with a specified accuracy of changes in time intervals. In relation to a black hole, our assessment confirms Barrow's conclusion \cite{Barrow_2020}: an increase in the degree of fractality reduces the time of evaporation of a black hole. The influence of quantum fluctuations on the information processing process is briefly considered.

\bibliographystyle{unsrt}
\bibliography{bibli2}

\begin{thebibliography}{10}

\bibitem{divakaran2024matter}
P.~P. Divakaran.
\newblock Matter in discrete space-times.
\newblock {\em arXiv preprint arXiv:2404.04548}, 2024.

\bibitem{Hossenfelder_2013}
Sabine Hossenfelder.
\newblock Minimal length scale scenarios for quantum gravity.
\newblock {\em Living Reviews in Relativity}, 16(1), January 2013.

\bibitem{Loll_1998}
Renate Loll.
\newblock Discrete approaches to quantum gravity in four dimensions.
\newblock {\em Living Reviews in Relativity}, 1(1), December 1998.

\bibitem{Williams_2006}
Ruth~M. Williams.
\newblock The generalized uncertainty principle and black hole remnants.
\newblock {\em Journal of Physics: Conference Series}, 33:38--48, 2006.

\bibitem{bolotin2020physics}
Yu.~L. Bolotin, A.~V. Tur, and V.~V. Yanovsky.
\newblock Physics of limit values at planck scale.
\newblock {\em arXiv preprint arXiv:2005.03984}.

\bibitem{mandelbrot1983fractal}
B.B. Mandelbrot.
\newblock {\em The Fractal Geometry of Nature}.
\newblock Einaudi paperbacks. Henry Holt and Company, 1983.

\bibitem{BTY16}
Yurii Bolotin, Anatoli Tur, and Vladimir Yanovsky.
\newblock {\em Chaos: Concepts, Control and Constructive Use}.
\newblock Understanding Complex Systems. Springer-Verlag Berlin Heidelberg,
  2017.

\bibitem{dewitt1964relativity}
C.~DeWitt and Bryce~S. DeWitt.
\newblock {\em Relativity, Groups and Topology (Houches Lecture)}.
\newblock Gordon and Breach Science Publishers, 1964.

\bibitem{Veneziano_1986}
G.~Veneziano.
\newblock A stringy nature needs just two constants.
\newblock {\em Europhysics Letters}, 2(3):199--204, 1986.

\bibitem{Amati_1988tn}
D.~Amati, M.~Ciafaloni, and G.~Veneziano.
\newblock Can space-time be probed below the string size?
\newblock {\em Phys. Lett. B}, 216:41--47, 1989.

\bibitem{Witten_1996}
E.~Witten.
\newblock Reflections on the fate of spacetime.
\newblock {\em Physics Today}, 49(4):24--31, 1996.

\bibitem{Adler_2001}
Ronald~J. Adler, Pisin Chen, and David~I. Santiago.
\newblock The generalized uncertainty principle and black hole remnants.
\newblock {\em General Relativity and Gravitation}, 33(12):2101–2108,
  December 2001.

\bibitem{Karolyhazy:1966zz}
F.~Karolyhazy.
\newblock {Gravitation and quantum mechanics of macroscopic objects}.
\newblock {\em Nuovo Cim. A}, 42:390--402, 1966.

\bibitem{hooft2009dimensional}
G.~'t~Hooft.
\newblock Dimensional reduction in quantum gravity.
\newblock {\em arXiv preprint gr-qc/9310026}, 2009.

\bibitem{Susskind_1995}
Leonard Susskind.
\newblock The world as a hologram.
\newblock {\em Journal of Mathematical Physics}, 36(11):6377–6396, November
  1995.

\bibitem{Ng_2003}
Y.~Jack NG.
\newblock Selected topics in planck-scale physics.
\newblock {\em Modern Physics Letters A}, 18(16):1073–1097, May 2003.

\bibitem{PhysRevD.83.084003}
Wayne~A. Christiansen, Y.~Jack Ng, David J.~E. Floyd, and Eric~S. Perlman.
\newblock Limits on spacetime foam.
\newblock {\em Phys. Rev. D}, 83:084003, Apr 2011.

\bibitem{Barrow_2020}
John~D. Barrow.
\newblock The area of a rough black hole.
\newblock {\em Physics Letters B}, 808:135643, September 2020.

\bibitem{Saridakis_2020b}
Emmanuel~N. Saridakis.
\newblock Modified cosmology through spacetime thermodynamics and barrow
  horizon entropy.
\newblock {\em Journal of Cosmology and Astroparticle Physics},
  2020(07):031–031, July 2020.

\bibitem{Saridakis_2020}
Emmanuel~N. Saridakis.
\newblock Barrow holographic dark energy.
\newblock {\em Physical Review D}, 102(12), December 2020.

\bibitem{Mamon_2021}
Abdulla~Al Mamon, Andronikos Paliathanasis, and Subhajit Saha.
\newblock Dynamics of an interacting barrow holographic dark energy model and
  its thermodynamic implications.
\newblock {\em The European Physical Journal Plus}, 136(1), January 2021.

\bibitem{adhikary2021barrow}
Priyanka Adhikary, Sudipta Das, Spyros Basilakos, and Emmanuel~N. Saridakis.
\newblock Barrow holographic dark energy in non-flat universe.
\newblock {\em arXiv preprint arXiv:2104.13118}, 2021.

\bibitem{SarBas2021}
Emmanuel~N. Saridakis and Spyros Basilakos.
\newblock The generalized second law of thermodynamics with barrow entropy.
\newblock {\em The European Physical Journal C}, 81:1434--6052, 2021.

\bibitem{Di_Gennaro_2022}
Sofia Di~Gennaro and Yen~Chin Ong.
\newblock Sign switching dark energy from a running barrow entropy.
\newblock {\em Universe}, 8(10):541, October 2022.

\bibitem{bolotin2024note}
Yu.~L. Bolotin and V.~V. Yanovsky.
\newblock Note on cosmographic approach to determining parameters of barrow
  entropic dark energy model.
\newblock {\em arXiv preprint arXiv:2402.1650}, 2024.

\bibitem{bolotin2024cosmology}
Yu.~L. Bolotin and V.~V. Yanovsky.
\newblock Cosmology based on entropy.
\newblock {\em arXiv preprint arXiv:2310.10144}, 2024.

\bibitem{BolotinCSW_2022}
Yu.~L. Bolotin, V.~A. Cherkaskiy, M.~I. Konchatnyi, Supriya Pan, and Weiqiang
  Yang.
\newblock Do current observations support transient acceleration of our
  universe?
\newblock {\em International Journal of Modern Physics D}, 31(05), March 2022.

\bibitem{PhysRevD123516}
Nobuyoshi Komatsu and Shigeo Kimura.
\newblock Entropic cosmology in a dissipative universe.
\newblock {\em Phys. Rev. D}, 90:123516, Dec 2014.

\bibitem{Komatsu_2014}
Nobuyoshi Komatsu and Shigeo Kimura.
\newblock Evolution of the universe in entropic cosmologies via different
  formulations.
\newblock {\em Physical Review D}, 89(12), June 2014.

\bibitem{Nojiri_2022n}
Shin’ichi Nojiri, Sergei~D. Odintsov, and Tanmoy Paul.
\newblock Early and late universe holographic cosmology from a new generalized
  entropy.
\newblock {\em Physics Letters B}, 831:137189, August 2022.

\bibitem{nojiri2023microscopic}
Shin'ichi Nojiri, Sergei~D. Odintsov, and Tanmoy Paul.
\newblock Microscopic interpretation of generalized entropy.
\newblock {\em arXiv preprint arXiv:2311.03848}, 2023.

\end{thebibliography}

\end{document}